\documentclass[journal]{IEEEtran}
\IEEEoverridecommandlockouts
\usepackage{amsmath}
\usepackage{color}
\usepackage{multicol}
\usepackage{array}
\usepackage{stfloats,balance}
\usepackage{amsthm} 
\usepackage{multirow}
\usepackage{amssymb}
\usepackage{float}
\usepackage{nccmath}
\usepackage{graphicx}
\usepackage{setspace}
\usepackage[algo2e, linesnumbered, ruled, vlined]{algorithm2e}
\usepackage{mathtools}
\usepackage{booktabs}

\usepackage{mwe}
\usepackage{cite}
\usepackage{amsmath,amssymb,amsfonts}
\usepackage{graphicx}
\usepackage{textcomp}
\usepackage{xcolor}
\usepackage{physics}
\usepackage{bm}
\usepackage[utf8]{inputenc} 
\usepackage[linesnumbered,ruled]{algorithm2e}
\usepackage{cite}
\usepackage{amsmath,amssymb,amsfonts,soul}
\usepackage{algorithmic}
\usepackage{graphicx}
\usepackage{textcomp}
\usepackage{xcolor}
\usepackage[normalem]{ulem}

\newcommand{\obs}[1]{\ensuremath{\langle #1 \rangle}}
\newcommand{\pure}[1]{\ensuremath{| #1 \rangle \langle #1 |}}

\newcommand{\BfPara}[1]{{\noindent\bf#1.}\xspace}

\def\BibTeX{{\rm B\kern-.05em{\sc i\kern-.025em b}\kern-.08em
    T\kern-.1667em\lower.7ex\hbox{E}\kern-.125emX}}
\begin{document}

\title{Projection Valued Measure-based Quantum Machine Learning for Multi-Class Classification}

\author{
    Won Joon Yun,
    Hankyul Baek,
    and
    Joongheon Kim,
    \IEEEmembership{Senior Member, IEEE}
    \thanks{This research was funded by the National Research Foundation of Korea (2022R1A2C2004869). \textit{(Corresponding author: Joongheon Kim).}}
    \thanks{Won Joon Yun, Hankyul Baek and Joongheon Kim are with the School of Electrical Engineering, Korea University, Seoul 02841, Republic of Korea (e-mails: \{ywjoon95,67back, joongheon\}@korea.ac.kr).}
}

\maketitle

\begin{abstract}
In recent years, quantum machine learning (QML) has been actively used for various tasks, e.g., classification, reinforcement learning, and adversarial learning. 
However, these QML studies are unable to carry out complex tasks due to scalability issues on input and output which is currently the biggest hurdle in QML. Therefore, the purpose of this paper is to overcome the problem of scalability. Motivated by this challenge, we focus on \emph{projection-valued measurements (PVM)} which utilizes the nature of probability amplitude in quantum statistical mechanics. By leveraging PVM, the output dimension is expanded from $q$, which is the number of qubits, to $2^q$. 
We propose a novel QML framework that utilizes PVM for multi-class classification. Our framework is proven to outperform \textit{the state-of-the-art (SOTA)} methodologies with various datasets, assuming no more than 6 qubits are used. Furthermore, our PVM-based QML shows about $42.2\%$ better performance than the SOTA framework.
\end{abstract}


\section{Introduction}\label{sec:1} 
IBM Quantum has publicized the 2022 development roadmap, announcing that 10--100k qubits quantum computers will be developed by 2026~\cite{roadmap2022}. Spurred by the recent advance of quantum computers, theoretical quantum supremacy has been corroborated through several simulations~\cite{arute2019quantum}. Due to its strengths, quantum computing is a strong candidate to replace classical computing in the near future. This has motivated many researchers to attempt combining existing fields with quantum computing~\cite{smc1,smc2}. Quantum machine learning (QML) is one of such emerging fields which exploits the strengths of both quantum computing and machine learning~\cite{biamonte2017quantum,xiong2022spl,du2015spl}. 
Basically, QML re-implements the existing machine learning (ML) solutions and neural networks (NN) using quantum neural networks (QNN)~\cite{pieee202105park,cong2019}, \textit{e.g.,} image classification~\cite{baek2022scalable,nikoloska2022}, reinforcement learning~\cite{yun2022quantum,icdcs2022yun}, federated learning (FL)~\cite{yun2022slimmable}, and communication networks~\cite{tetc1}. 
Compared to ML, QML is still in extremely early stages of research, with many terminologies and theories not yet clearly defined~\cite{kwak2021quantum}. In addition, it also has important challenges to resolve such as scalability, trainability, and difficulty in large-scale simulations~\cite{cerezo2022challenges}.
Quantum convolution neural network (QCNN) is one of the innovative solutions which has resolved trainability~\cite{PhysRevX.11.041011} and scalability on input data~\cite{baek2022scalable}.

However, QML still suffers from scalability on outputs. Thus, QML could only be used for simple tasks (\textit{e.g.}, binary classification~\cite{chen2021federated,chehimi2022quantum,ref1,ref3,ref4}), or classical NNs must be used at the end of QNN in order to obtain good performance~(\textit{i.e.}, quantum-classical hybrid computing~\cite{baek2022sqcnn3d,ref2,ref5}). Despite the importance of pure QML in proving quantum supremacy, the multi-class classification with a pure quantum version has not been considered yet.
\begin{figure}[t!]
\centering
\includegraphics[width=0.95\columnwidth]{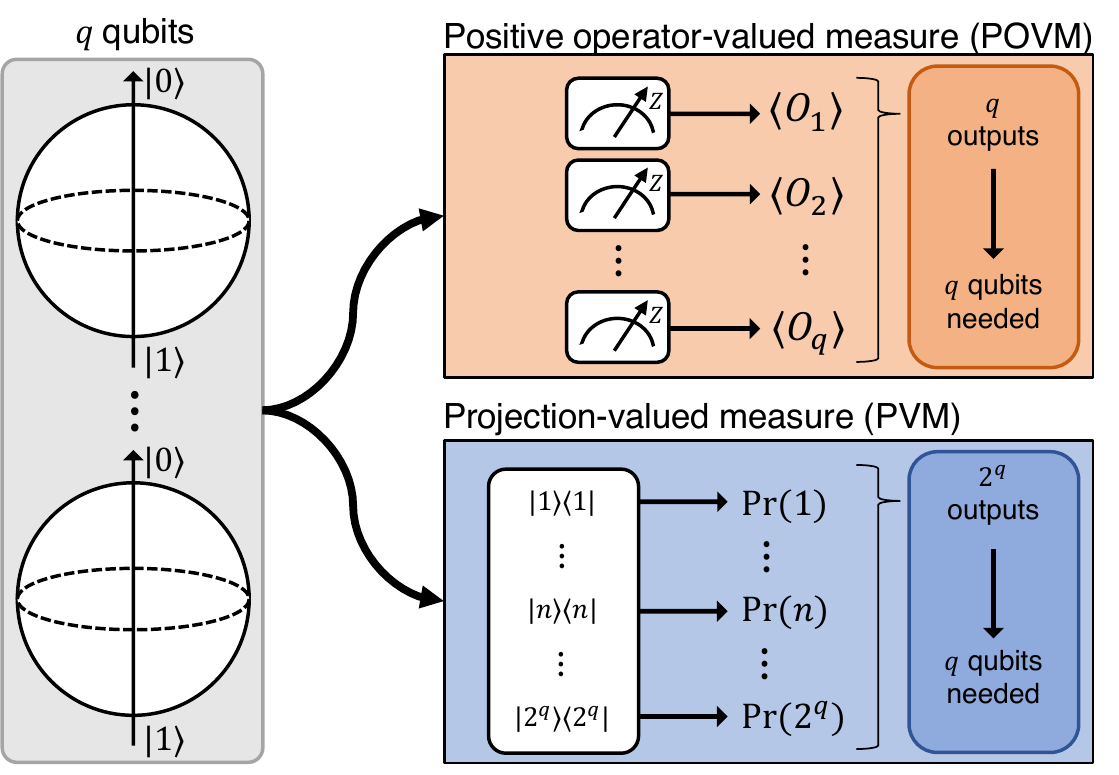}
\caption{The difference between POVM and PVM. In this paper, we utilize POVM for image processing as well as the PVM method for multi-class prediction.}
\end{figure}

Motivated by these trends, we propose the methodology of QML for multi-class classification. To understand how QML can achieve multi-class classification, we investigate quantum measurement theory and its applications. Most quantum computers and QML frameworks follow the \textit{positive operator-valued measure (POVM)} to obtain the output~\cite{banchi2021generalization}. However, the size of the POVM-based output in $q$-qubit system cannot exceed the number of qubits, \textit{i.e.}, $q$ and this hinders the output scalability of QNN.
Next, we focus on the \textit{projection-valued measure (PVM)}, which is a special form of POVM~\cite{jacobs2014quantum}. By utilizing PVM in QNN, $2^q$ outputs (named observables) can be produced by utilizing whole bases of $q$-qubits. By leveraging this nature of PVM, \textit{the quantum machine learning framework for multi-class classification} without any classical NN methodologies is proposed. In addition, we propose a probability amplitude regularizer to normalize the observables with respect to true labels such that the performance of QML framework can be significantly improved. 

\BfPara{Contributions} The contributions of our proposed QML framework are listed as follows.
\begin{itemize}
\item We first propose a novel PVM-based QNN for multi-class classification which does not use any classical NN methodologies. Our proposed QML framework provides an insightful bridge between quantum computing and machine learning. 
\item Moreover, we show the quantum supremacy in QCNN computation operations by comparing the computational complexity of our proposed model to that of classical NN methodologies.
\item Finally, we improve the scalability of QML without classical NN methodologies by increasing the input up to $32\times 32\times 3$ and output up to 26 classes, respectively.
\end{itemize}

\BfPara{Organization} 
The rest of this paper is organized as follows.
Sec.~\ref{sec:2} introduces the preliminary knowledge of QML. 
After that, Sec.~\ref{sec:3} presents the proposed novel QML framework for multi-class classification; and then, 
Sec.~\ref{sec:4} evaluates the performance of the proposed framework. 
Lastly, Sec.~\ref{sec:5} concludes this paper and presents future research directions. 

\section{Preliminaries: Quantum Machine Learning}\label{sec:2}
\subsection{Notation and Mathematical Setup}\label{sec:2-1}
In this paper, we use notations  $\Theta = \{\bm  \theta^{\text{enc}}; \bm \theta^{\text{PQC}}\}$ for trainable parameters.
We define $\zeta \triangleq \{(\mathbf{X}, \bm y)\}$ as a sampled mini-batch where $\mathbf{X}$ and $\bm y$ stand for the sampled input data and the corresponding labels from mini-batch, respectively. The labels $\bm y \triangleq \{y_n\}_{n=1}^{|\bm y|}$ are one-hot vectors where $y_n = 1$ if the true labels are $n \in \mathbb{N}[1, |\bm y|]$, and the other elements are $0$. 
The extracted features are denoted as $\hat{\mathbf{X}}$. 
Dirac-notation is used to represent the quantum state and its operations. In addition, the operators $(\cdot)^\dag$ and $\otimes$ stand for complex conjugate transpose and tensor product, respectively. 
We use the two terms $Q$ and $q$ separately to indicate a $Q$-qubit system that encodes classical data and a $q$-qubit system that performs prediction.
The $Q$ qubit quantum state is defined as follows,
\begin{equation}
    \ket{\psi} = \sum^{2^{Q}}_{n=1} \alpha_n \ket{n},
\end{equation}
where $\alpha_n$ and $\ket{n}$ stand for the probability amplitude and $n$-th basis in Hilbert space, respectively. By the definition of Hilbert space (i.e., $\mathcal{H}^{\otimes Q} \equiv \mathbb{C}^{2^Q}$), probability amplitude  $\forall \alpha_n \in \mathbb{C}$ satisfies the following equation, \textit{i.e.,} $\sum^{2^Q}_{n=1} |\alpha_n|^2 = 1$.

\begin{figure*}[t!]
    \centering
    \includegraphics[width=\linewidth]{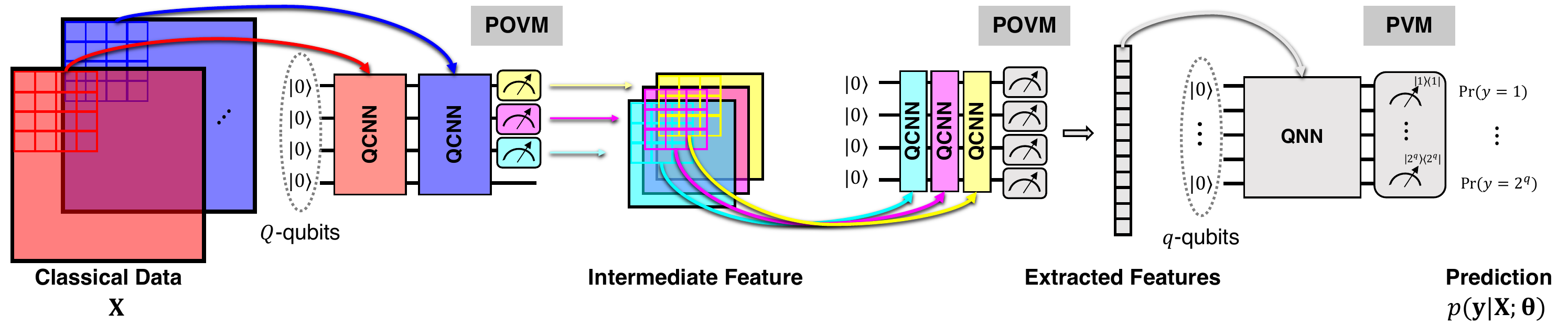}
    \caption{The computing processes on QCNN and QNN. Each color (e.g., \textit{red}, \textit{blue}, \textit{yellow}, \textit{magenta}, or \textit{cyan}) indicates the channel. Each 2D grid indicates the patch in each channel. When the quantum state is measured in QCNN, the expected value of projection for each qubit represents the value of the output channel. The extracted features are vectorized classical data, which is given as an input of QNN. QNN makes a probability measure that corresponds to the prediction of classes.}
    \label{fig:process}
\end{figure*}

\subsection{Quantum Measurement Theory for Classification}\label{sec:2-2}
In measurement theory, two main streams of study exist in the literature, \textit{i.e.}, \textit{POVM}, and \textit{PVM} \cite{jacobs2014quantum}. 
POVM generates outputs by projecting quantum states into Pauli-$Z$ measure devices for every qubit, where Pauli-$Z$ is $\mathbf{Z} = \text{diag}(1, -1)$. The $n$-th projection matrix is defined as 
$\mathbf{P}_n \triangleq \mathbf{I}^{\otimes n-1}\otimes \mathbf{Z} \otimes \mathbf{I}^{\otimes Q-n}$, where $\mathbf{I}$ stands for $2\times2$ identity matrix, respectively. 
PVM calculates the expectation value of a projection $\mathbf{P}_n$ as random variable denoted as $O_n \in \mathbb{R}[-1, 1]$ where $\forall n \in \mathbb{N}[1,Q]$. 
Its expectation value of $m$-th qubit is $
\obs{O_n} = \bra{\psi} \mathbf{P}_n \ket{\psi}$. According to \cite{jacobs2014quantum}, the expectation value of a projection is used with the softmax function and temperature parameter $\beta \in \mathbb{R}$ to make a prediction for $q$-classes as follows,
\begin{equation}
\Pr\nolimits_{POVM}(y=n) = \frac{\exp(\beta \obs{O_n})}{\sum^q_{n=1}\exp(\beta \obs{O_n})}. \label{eq:povm}
\end{equation}

Note that the number of qubits $q$ should satisfy $q \geq |\bm y|$ in POVM.
Additionally, POVM with classical NNs is widely used to make large-scale predictions as it is not limited by the condition $q \geq |\bm y|$. In this paper, PVM is considered to be a special type of POVM and when the probability is measured by projecting the quantum state into projector $\{\pure{n}\}_{n=1}^{2^q}$, the output is expressed as follows,
\begin{equation}
\Pr\nolimits_{PVM}(y=n) = \langle \psi| n \rangle  \langle n |\psi \rangle =  |\langle \psi| n \rangle |^2 = |\alpha_n|^2. \label{eq:pvm}
\end{equation}

Note that the output of PVM is a probability distribution which can be represented as
\begin{equation}
\Pr\nolimits_{PVM}(y=n)  = \sum^{2^{q}}_{n=1} |\alpha_n|^2 = 1. 
\end{equation}

In this paper, both POVM and PVM are used for QCNN-based image processing and QNN-based classification.

\section{Quantum Machine Learning Framework for Multi-Class Classification}\label{sec:3}
\subsection{Architecture}
Our QML framework for multi-class classification consists of QCNNs and QNN. The structure of a QCNN/QNN is tripartite: the state encoding, linear transform with parametrized quantum circuits (PQCs), and the measurement~\cite{killoran2019continuous}. 
First of all, we consider data-reuploading \cite{P_rez_Salinas_2020} for state encoder because the classical input size is larger than the number of qubits. To successfully encode the classical input data $\mathbf{X}$, the $\mathbf{X}$ is split into $[\mathbf{x}_1; \cdots ; \mathbf{x}_{c_{\text{in}}}]$ which are encoded to probability amplitudes and the encoding process is expressed as follows,
\begin{equation}
|\psi_{\text{enc}}\rangle  = U(\bm {\theta}_{c_{\text{in}}})U(\mathbf {x}_{c_{\text{in}}})\cdots U(\bm {\theta}_1)U(\mathbf{x}_1) |\psi_0\rangle,  \label{eq:data-reuploading}   
\end{equation}
where $|\psi_0\rangle$ denotes the initial quantum state, \textit{e.g.}, the first standard basis of $2^Q$-dimensional vector and $\forall \bm \theta_c \subset \bm \theta_{\text{enc}}$, and $\forall c \in \mathbb{N}[1,c_{\text{in}}]$. The encoded state $|\psi_{\text{enc}}\rangle$ is processed with PQCs where the result is presented as $|\psi_{\text{PQC}}\rangle = U(\bm \theta^{\text{PQC}})|\psi_{\text{enc}}\rangle$. The processed quantum state $|\psi_{\text{PQC}}\rangle$ is measured by POVM in QCNN, and PVM in QNN, respectively.
\subsection{Pipeline}
\BfPara{POVM-based QCNN for Image Processing} As presented in Fig.~\ref{fig:process}, QCNN takes input from classical data and returns the extracted feature. Algorithm~\ref{alg:local}(lines 2--13) presents the progression of  image processing with QCNN. 
In QCNN, the input has $W \times H \times c_{\text{in}}$ shape. Note that $\hat{W}$ and $\hat{H}$ depend on the kernel size $\kappa$, stride $s$, and padding $d$, \textit{i.e.,} $\hat{W} = ({W + d})/{s}$ and $\hat{H} = ({H + d})/{s}$. Each patch of input is feed-forwarded to QCNN by data-reuploading method \eqref{eq:data-reuploading}. Then, the pooling of QCNN is carried out by POVM. The $n$-th expected value of POVM $\obs{O_n}$ corresponds to a scalar value of $n$-th channel and the output has $\hat{W} \times \hat{H} \times c_{\text{out}}$ shape. The computational complexity of our QNN per layer is $\mathcal{O}(\hat{W} \cdot \hat{H} \cdot \kappa^2  \cdot c_{\text{in}})$, where $\kappa$ denotes the kernel size, whereas the computational complexity of classical CNN is $\mathcal{O}(\hat{W} \cdot \hat{H} \cdot \kappa^2 \cdot c_{\text{in}} \cdot c_{\text{out}})$. Note that the input and output channels in QCNN are scalable under the constraint $\forall c_{\text{in}}, c_{\text{out}} \in \mathbb{N}[1, Q]$.

\begin{figure}[t!]
\begin{tabular}{@{}c@{}}
    \includegraphics[width=\columnwidth]{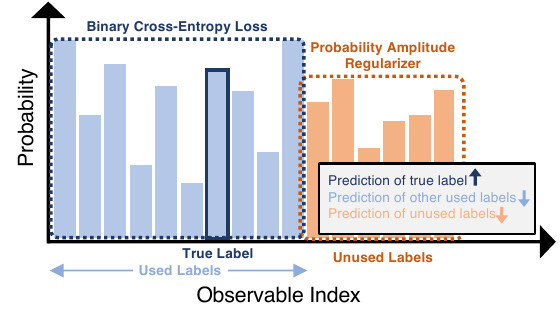}
    \end{tabular}
    \caption{The illustration of the objective function ($|\bm y | = 10$ and $q = 4$). Two loss functions are used in this paper (\textit{i.e.,} binary cross-entropy loss and probability amplitude regularizer). Leveraging our proposed loss function has increased the prediction frequency of true label while decreasing the prediction frequency of other used labels, and unused labels.}
    \label{fig:loss}
\end{figure}

\BfPara{PVM-based QNN for Classification}
Generally for quantum circuits, the number of classes $|\mathbf{y}|$ is larger than the number of qubits $q$. However, it is hard to train QCNN/QNN when the number of qubits is increased. Thus, softmax-based POVM such as \eqref{eq:povm} is limited to simple multi-class classification (\textit{e.g.}, binary classification or 4 classes classification). Thus, in order to extend quantum computing to complex multi-class classification, we consider the PVM-based QNN, where its process is presented in Algorithm~\ref{alg:local}(lines 14--20).
Similar to QCNN processing, the extracted features are encoded with \eqref{eq:data-reuploading}.
To enable multi-class classification, we design the measurement of QNN with PVM. Then, the probability of $2^q$ observables can be obtained when a quantum state $\ket{\psi}$ is projected into pure density matrices, \textit{i.e.}, $\{\pure{1}, \cdots, \pure{n}, \cdots,\pure{2^q}\}$. The probability of $n$-th class is expressed as follows,
\begin{equation}
p_n = \Pr(y=n|\mathbf{X};\Theta) = |\alpha_n|^2.  
\end{equation}

Note that our framework does not require softmax function to make predictions with logits since the probability amplitude $\alpha_n$ is mapped to the probability of class $n$ directly. Additionally, the softmax temperature coefficient is not required as well~\cite{jerbi2021variational}. However, with a na\"{i}ve consideration, PVM suffers from increased error probability due to the presence of unused classes. We will elaborate on how to resolve it on next. 
\begin{algorithm2e}[t]
    \SetCustomAlgoRuledWidth{0.44\textwidth}  
\caption{\normalsize QML for Multi-Class Classification}
\label{alg:local} 
\textbf{Notation.} Processing data $\breve{\mathbf{X}}$, Processed data $\tilde{\mathbf{X}}$, stride $s$, kernel size $\kappa$, and the number of QNN depths $D$\;
\textbf{Image Processing}($\mathbf{X}$):\\
\Indp $\breve{\mathbf{X}} \leftarrow \mathbf{X}$\;

    \For{$w \in \{0, s, 2s, \cdots,W - s\}$}
     {
         \For{$h \in \{ 0, s, 2s, \cdots, H - s\}$}
            {    
                Initialize quantum state, $\ket{\psi} \leftarrow \ket{0}$\;
            \For{$c \in \{1,2, \cdots,  c_{\text{in}}\}$}
                {
                Prepare data, $\breve{\mathbf{x}}\leftarrow \breve{\mathbf{X}}_{w:w+\kappa,h:h+\kappa, c}$\;
                Upload data, $\ket{\psi} \leftarrow U(\bm {\theta}_c)\cdot U(\breve{\mathbf{x}})\cdot \ket{\psi}$\;
                }
                \For{$c \in \{1,2, \cdots,  c_{\text{out}}\}$}
                {
                Measure through POVM and collects outputs, $\tilde{\mathbf{X}}_{\frac{w}{s}, \frac{h}{s}, c}\leftarrow \obs{O_c}_{\breve{\mathbf{x}},\Theta}$\;
                }
             }
     }
     Get intermediate feature, $\breve{\mathbf{X}} \leftarrow \tilde{\mathbf{X}}$\;
  \textbf{Output:} Extracted feature $\mathbf{\hat{X}}$\;
\Indm \textbf{Prediction}($\mathbf{\hat{X}}$):\\
\Indp Initialize $\mathbf{p} \leftarrow \emptyset$\;
Prepare the pure density matrices, $\{\pure{1}, \cdots, \pure{n}, \cdots \pure{2^q}\}$\;
\For{$m \in\{1,\cdots,2^q\}$}
{Obtain prediction value of class $c$\;
 $\mathbf{p}\leftarrow \mathbf{p} \cup p_c$\;}
\textbf{Output:} Prediction $\mathbf{p}$\;

\end{algorithm2e}

\subsection{Training}
In this paper, we design a new regularizer to reduce performance degradation incurred by training errors due to the probability of unused classes. In Fig.~\ref{fig:loss}, we compose the objective function with binary cross-entropy function $\mathcal{L}_{\text{BCE}}$ and \textit{probability amplitude regularizer} $\mathcal{L}_{\text{PAR}}$ for probability regarding unused class indices, which is expressed as follows,
\begin{align} \label{eq:BCE}
\mathcal{L}_{\text{BCE}}(\Theta;\mathbf{X}) &= - \sum^{|\bm y|}_{n=1}[y_c \log p_c + (1-y_c) \log (1-p_c)], \\
\mathcal{L}_{\text{PAR}}(\Theta;\mathbf{X}) &=-\sum^{2^q}_{n'>|\bm y|}\log (1-p_{n'}), \label{eq:PAR}
\end{align}
where $q\geq \lceil \log_2(|\bm y|) \rceil$. 
Thus, we finalize the train loss function for a one-step single update as follows,
\begin{equation}
\mathcal{L}(\Theta; \zeta) =  \frac{1}{|\zeta|}\sum_{(\mathbf X, \bm y) \in {\zeta}} [\mathcal{L}_{\text{BCE}}(\Theta;\mathbf{X}) + \mathcal{L}_{\text{PAR}}(\Theta;\mathbf{X})].
\end{equation}

Next, the gradient of the train loss function is obtained as follows. 
Because a quantum computer cannot use classical training methods, \textit{e.g.}, back-propagation with chain rule, the $0$-th order optimization is utilized to estimate the gradient, called \textit{parameter-shift rule}~\cite{Chen_2017,crooks2019gradients}. By calculating the symmetric difference quotient of loss $\mathcal{L}$, the loss gradient is obtained as follows, 
\begin{align}
\frac{\partial \mathcal{L}(\Theta;\zeta)}{\partial \theta_{m}} &= \frac{\partial \mathcal{L}(\Theta;\zeta)}{\partial f(\Theta;\zeta)}
\cdot \frac{\partial f(\Theta;\zeta)}{\partial \theta_{m}} ,\\
\text{s.t.~~} \frac{\partial f(\Theta;\zeta)}{\partial \theta_{m}} & = f\Big(\Theta + \frac{\pi}{2} \mathbf{e}_{m};\zeta\Big) - f\Big(\Theta - \frac{\pi}{2} \mathbf{e}_{m};\zeta\Big), \label{eq:ps-rule}
\end{align}
 where $f(\Theta;\zeta)$ denote the output of QNN.
 In addition, $\mathbf{e}_{m}$ is an one-hot vector that eliminate all components except $\theta_{m}$, \textit{i.e.,} $\Theta \cdot \mathbf{e}_{m} = \theta_{m}$ and $\forall m \in \mathbb{N}[1, |\Theta|]$, respectively.  Finally, the trainable parameters in our QML framework are updated as $\Theta \leftarrow \Theta - \eta \nabla_\Theta \mathcal{L}(\Theta)$, where $\eta$ is a learning rate.

\section{Performance Evaluation}\label{sec:4}

\begin{table}
\caption{Experimental Settings}
\centering\small
\begin{tabular}{@{}cccc@{}}
\toprule[1.5pt]
\textbf{Dataset} & \textbf{Input size}  & \textbf{\# of classes} & \textbf{\# of qubits ($q$)}\\\midrule[1pt]
MNIST & $28\times28\times 1$ & 10 & 4\\
FashionMNIST & $28\times28\times 1$ & 10 & 4\\
CIFAR10 & $32\times32\times 3$ & 10 & 4\\
EMNIST-letters & $28\times28\times1$ & 26 & 6\\
\bottomrule[1.5pt]
\end{tabular}
\label{tab:dataset}
\end{table}\begin{figure*}[t!]\centering
\small
\begin{tabular}{@{}c@{}c@{}c@{}c@{}}
\includegraphics[width=.25\linewidth]{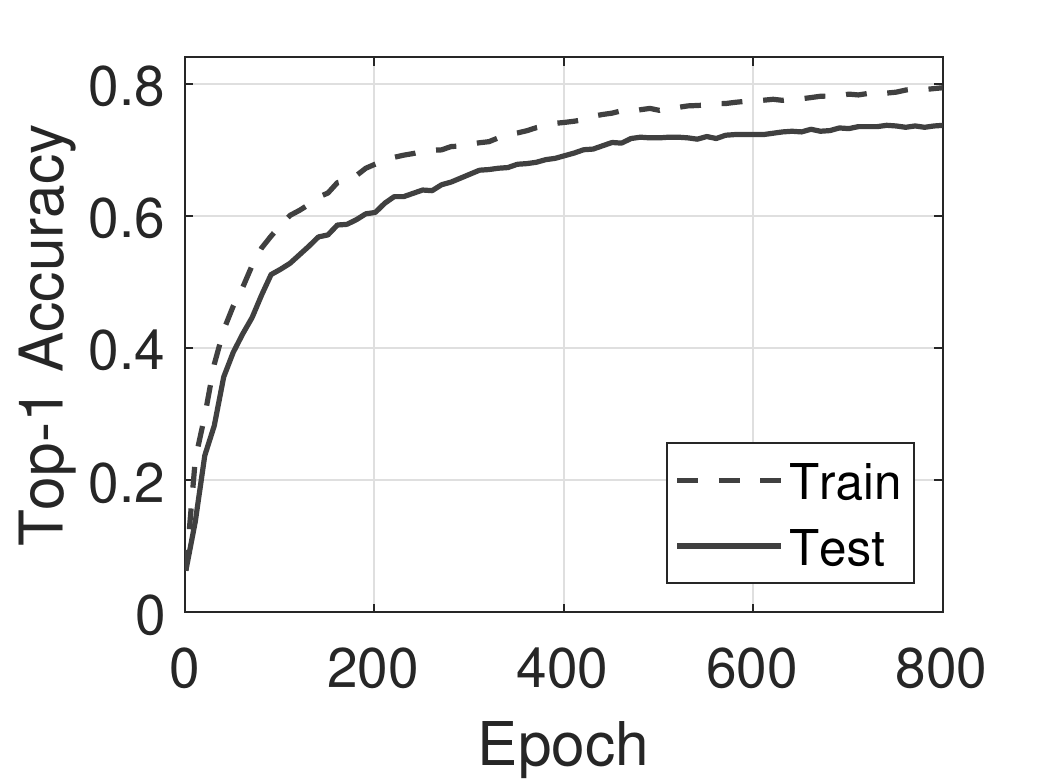} &
\includegraphics[width=.25\linewidth]{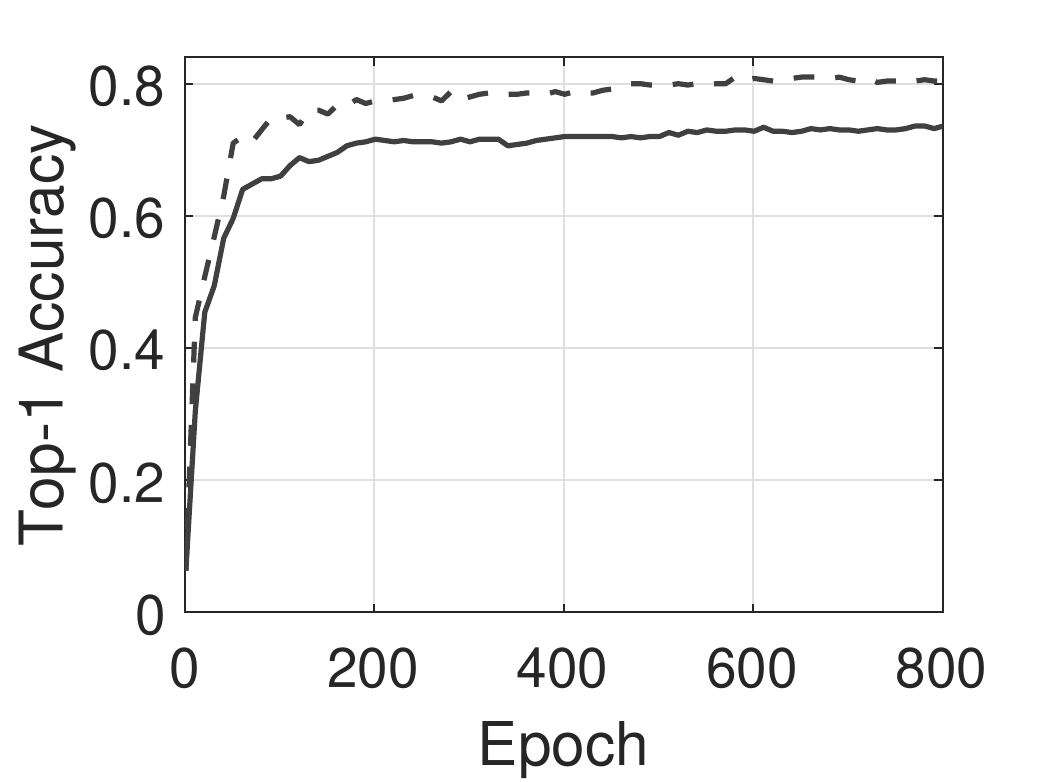} & 
\includegraphics[width=.25\linewidth]{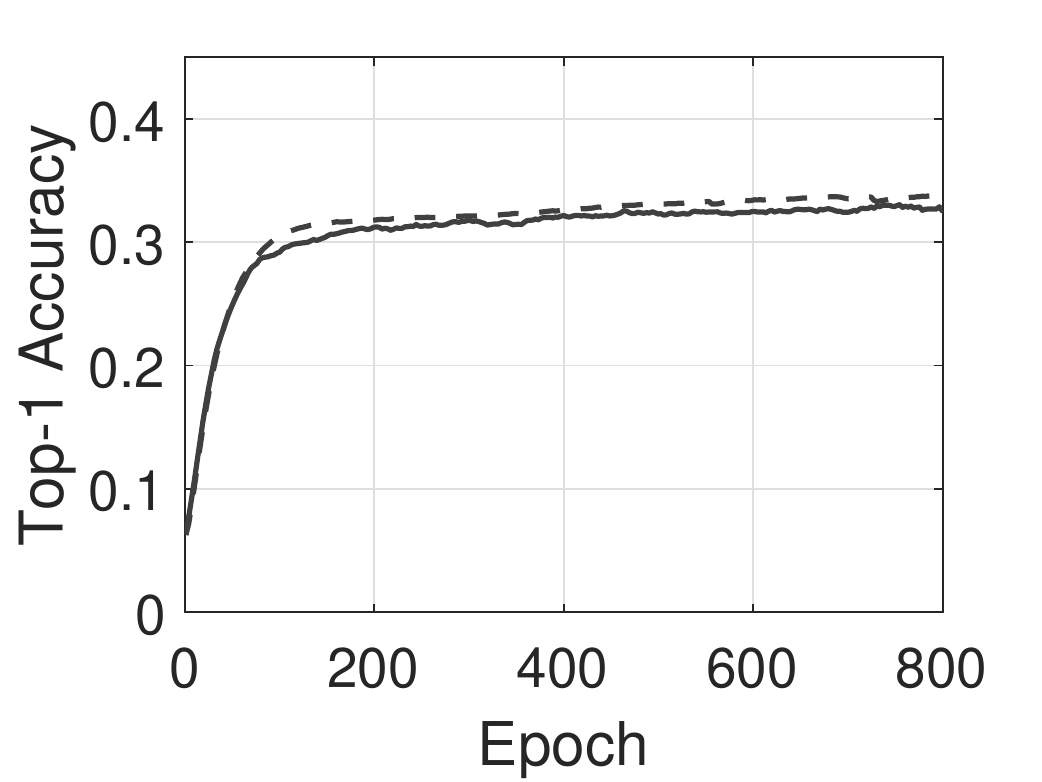} & 
\includegraphics[width=.25\linewidth]{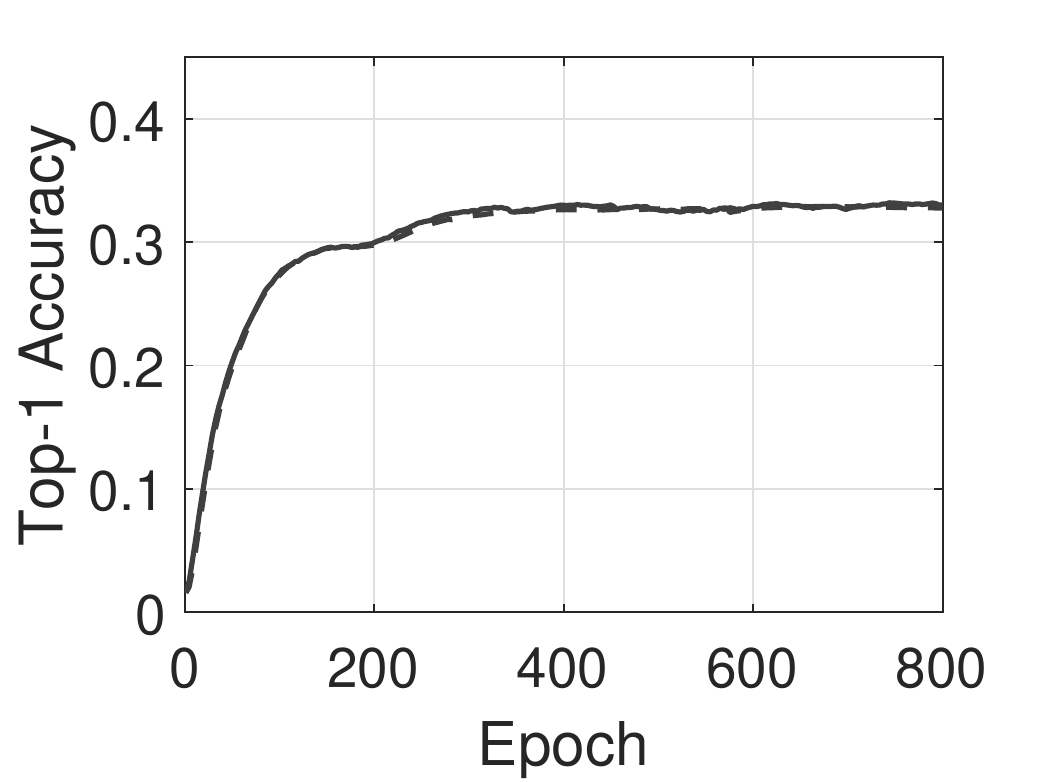} \\
(a) MNIST& (b) FashionMNIST & (c) CIFAR10 & (d) EMNIST-letters 
\end{tabular}
\caption{Learning curve with various datasets.} \label{fig:5}
\end{figure*}
\begin{figure}[t!]
\centering
\begin{tabular}{@{}c@{}c@{}}
\includegraphics[width=.5\columnwidth]{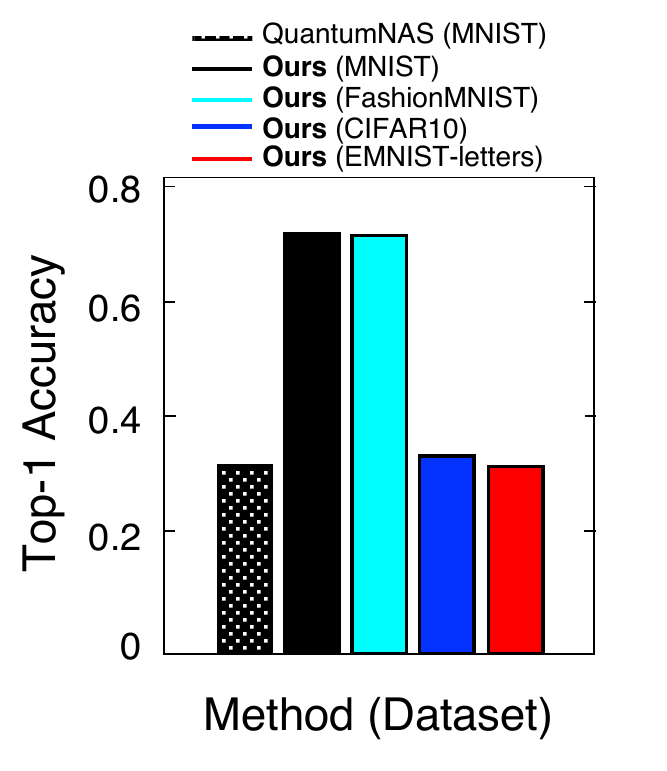} &
\includegraphics[width=.5\columnwidth]{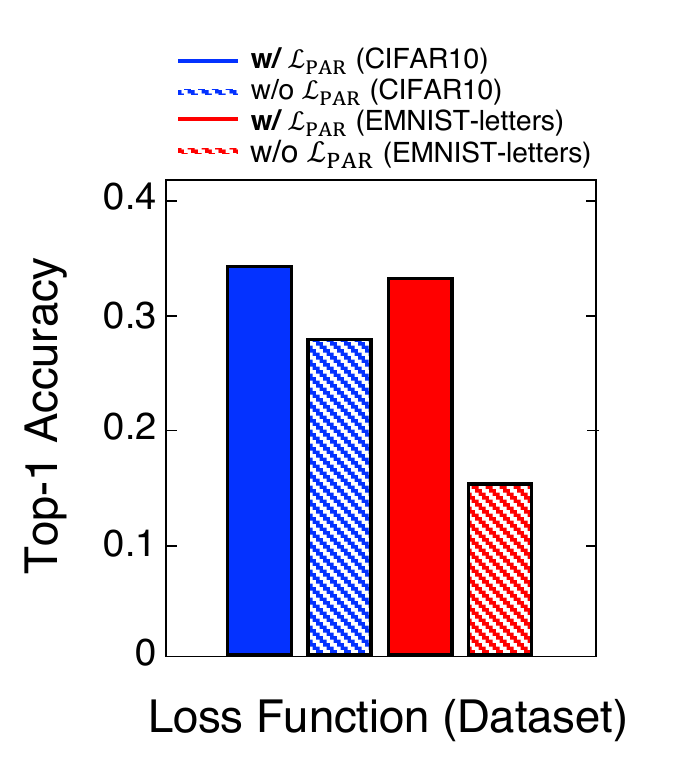}\\
\small (a) Comparison with SOTA \cite{quantumnas} & 
\small (b) Impact of Regularizer
\end{tabular}
\caption{Additional experiments: (a) shows the comparison between our framework and QuantumNAS \cite{quantumnas}, and (b) shows the ablation study of probability amplitude regularizer.}
\label{fig:6}
\end{figure}
\subsection{Experiment Setup}
In this section, we mainly investigate the trainability of our QML framework with various datasets and the impact of a probability amplitude regularizer via performance evaluation. For this, we benchmark the training accuracy, test accuracy, comparison to \textit{state-of-the-art (SOTA)}, and the ablation of the regularizer. 
Because our main scope is to investigate the scalability of QML, we do not interpolate the input data size nor use selected classes. The datasets used in this paper are listed in Table \ref{tab:dataset}. 
We use two QCNN models for image processing where $4\times4$ kernel is commonly used with stride $s=3$ and the kernels have $3$ channels. Our framework utilizes $4$ qubits and controlled unitary gates for all quantum circuits, \textit{e.g.}, QCNN, and QNN~\cite{sleator1995realizable}. 
The hyperparameters used in this paper are as follows, \textit{i.e.}, Adam optimizer, $8\times10^{-3}$ for initial learning rate, and $1$,$024$ for batch size. Note that \typeout{we do not consider quantum noise and }all experiments for the proposed QML framework are conducted in classical computers with Python v3.8.10 and \texttt{torchquantum}~\cite{quantumnas}.

\subsection{Numerical Results}
\BfPara{Trainability} 
Fig.~\ref{fig:5} shows the learning curves with various datasets. Our framework converges to sub-optimal for all datasets. In Fig.~\ref{fig:5}(a), the test accuracy increases from $6.25$\% to $74.2$\%, and the domain gap between the train set and the test set is about $5.7$\% between the top-1 accuracies. 
In Fig.~\ref{fig:5}(b), the test accuracy increases from $6.25$\% to $73.8$\%, where $6.8$\% gap exists. In Fig.~\ref{fig:5}(c)/(d), the top-1 accuracy for CIFAR10 and EMNIST-letters show $34.4$\% and $33.1$\%, respectively, and the domain gap between the train set and test set is lower than that of MNIST and FashionMNIST.

\BfPara{Comparison to SOTA} 
We compare our framework to QuantumNAS~\cite{quantumnas}. According to \cite{quantumnas}, the input is $6 \times 6$ sized MNIST images, and QuantumNAS considers quantum noise to exist on the quantum devices being used. However, we benchmark our framework with the original size of the dataset as shown in Table \ref{tab:dataset} without considering quantum noise. 
As shown in Fig.~\ref{fig:6}(a), our framework outperforms QuantumNAS by $44.2$\% regarding top-1 accuracy given the MNIST dataset. Moreover, our framework is benchmarked with various datasets, \textit{i.e.,} FashionMNIST, CIFAR10, EMNIST-letters using less than $7$ qubits, whereas QuantumNAS uses $10$ qubits. 
In summary, PVM enables not only multi-class classification but also achieves higher accuracy than SOTA.

\BfPara{Ablation Study of Regularizer} In this paper, we propose a \textit{probability amplitude regularizer} and conduct an ablation study of the regularizer with two datasets (i.e., CIFAR10, EMNIST-letters). As shown in Fig.~\ref{fig:6}(b), the top-1 accuracy of the training scheme with $\mathcal{L}_{\text{PAR}}$ shows better performance of 7.5\% for CIFAR10 and 17.8\% for EMNIST-letters. It is because the prediction frequency of other used labels is decreased via the probability amplitude regularizer. Thus, the significant influence of our proposed regularizer in our QML framework for multi-class classification is demonstrated.

\section{Conclusions and Future Work}\label{sec:5}
We aim to expand the limited input and output dimensions of QML and to achieve multi-class classification by leveraging quantum circuits. In order to carry out multi-class classification, we utilize QCNN with POVM as well as QNN with PVM. Additionally, we propose a probability amplitude regularizer which fits the probability distribution of observables to the probability distribution of classification. Via extensive results showing the proposed model outperforming SOTA methodologies, we corroborate the efficacy of PVM-based QML for multi-class classification.

Our future work directions are as follows, \textit{i.e.}, (1) quantum reinforcement learning with the large action spaces or multi-agent setting with less qubits and (2) quantum object detection by leveraging POVM for bounding box prediction and PVM for classification. 

\bibliographystyle{IEEEtran}
\balance
\bibliography{reference}
\end{document}